# Hidden Populations in Software Engineering: Challenges, Lessons Learned, and Opportunities

Ronnie de Souza Santos
University of Calgary
Calgary, AB, Canada
ronnie.desouzasantos@ucalgary.ca

Kiev Gama
Universidade Federal de Pernambuco
Recife, PE, Brazil
kiev@cin.ufpe.br

## ABSTRACT

The growing emphasis on studying equity, diversity, and inclusion within software engineering has amplified the need to explore hidden populations within this field. Exploring hidden populations becomes important to obtain invaluable insights into the experiences, challenges, and perspectives of underrepresented groups in software engineering and, therefore, devise strategies to make the software industry more diverse. However, studying these hidden populations presents multifaceted challenges, including the complexities associated with identifying and engaging participants due to their marginalized status. In this paper, we discuss our experiences and lessons learned while conducting multiple studies involving hidden populations in software engineering. We emphasize the importance of recognizing and addressing these challenges within the software engineering research community to foster a more inclusive and comprehensive understanding of diverse populations of software professionals.

## 1 INTRODUCTION

Sampling is defined as selecting a subset (or sample) of individuals from a larger population to collect data and make inferences about that population [25]. In the context of Software Engineering, sampling focuses on selecting specific individuals—such as software engineers—from a larger pool of professionals to gain insights into the collective workforce. This strategy allows us to concentrate on a group of professionals to research their behaviors, practices, and experiences within the software development process [4, 15]. Thus, when sampling software professionals, it is crucial to select a subset that authentically mirrors the diverse array of backgrounds, skills, and experiences existing within the software industry.

From its methodological roots, the sampling process must allow statistical generalizations to the broader population (positivist research) and allow analytical generalization to theories or transferability to other contexts (constructivist research) [4, 10, 15]. In theory, this means that the conclusions derived from a sample of software professionals should allow researchers, practitioners, and stakeholders to derive knowledge about the specific aspects investigated within that sample, such as software development practices. Hence, an effective sampling method should provide insights about the sampled individuals and allow for meaningful extrapolation of the data to a wider software engineering domain [25, 40].

In practical terms, sampling for generalization can be challenging, particularly in certain software engineering contexts. Previous research has highlighted the scarcity of random sampling in empirical software engineering (ESE) studies due to the lack of available reliable sampling frames (e.g., population lists) [2, 4]. This challenge becomes more apparent in ongoing research on diversity in software engineering, as conventional sampling methods may not always be feasible when studying certain underrepresented groups of software professionals.

The study of underrepresented groups in software engineering brings forward the concept of hidden populations. Hidden populations refer to groups or individuals within a larger population that is challenging to identify, access, or study due to their secretive, marginalized, or elusive nature [1, 22]. Common examples in broader literature include homeless individuals, people with stigmatized medical conditions, those who have faced violence and trauma, and several other marginalized groups in our society [27, 36, 43]. Based on this definition, certain groups of software professionals can be classified as hidden populations, including female software professionals facing harassment in male-majority teams, LGBTQIA+ individuals in software engineering, indigenous software professionals, and other minority groups working in the field.

Considering the broad challenges in sampling—such as non-representative samples, inadequate sample sizes, and sampling bias—as well as specific software engineering issues—such as generalizing from students to software professionals or from open-source to closed-source projects—studying hidden populations in software engineering poses additional challenges [4]. Primarily, challenges emerge from identifying participants from some underrepresented groups for research and reporting findings derived from small sample sizes (either in a constructivist or positivist approach). In this sense, this study delves into these issues as we share insights from our experience investigating hidden populations of software professionals.

From this introduction, our paper is organized as follows. Section 2 explores studies involving hidden populations in software engineering. In Section 3, we outline experiences in researching hidden populations, presenting faced challenges, discussing lessons learned, and introducing opportunities within this context. Lastly, Section 4 is focused on our general conclusions.

## 2 HIDDEN POPULATIONS

A hidden population represents a group that lacks a clear, accessible sampling framework for research or study purposes [1, 22]. One of the characteristics associated with this type of population is that acknowledging belonging to this group may pose risks or threats to the individuals, usually due to societal stigma, legal implications, or the sensitive nature of their characteristics or activities [22]. Consequently, researchers encounter challenges in effectively identifying, studying, and supporting this population [6].

Hidden populations present a challenge for outsiders to access their individuals, prompting researchers to adapt sampling methods when studying these groups [44]. For known hidden populations, methods like snowballing and purposive sampling are often recommended [3, 42]. However, identifying unknown or truly invisible groups lacks sufficient support within research methodologies [1]. In any of the cases, probability sampling becomes challenging, impeding result generalization across these populations, regardless of the approach used [1, 3, 22, 31, 37, 42].

Hidden populations encompass a range of marginalized groups, comprising homeless individuals, undocumented immigrants, and sex workers, as well as individuals engaged in illicit activities such as drug use. Moreover, members of the LGBTQIA+ community facing discrimination, individuals with stigmatized health conditions like mental illnesses or HIV, and those enduring physical or psychological violence, including harassment, often remain secluded due to societal constraints, fear, or coercion. Consequently, these groups face challenges in being reached for research studies and in accessing essential support services [27, 36, 43].

In software engineering, hidden populations manifest in various forms, including groups that align with the broader definition found in the literature. For instance, LGBTQIA+ software professionals, women experiencing harassment within software development environments, and software professionals contending with mental health conditions such as depression. Within the software engineering literature, some publications addressed hidden populations, including experienced green software practitioners [30], LGBTQIA+ software professionals [12, 13, 18], LGBTQIA+ software engineering students [34], cyber-physical systems engineers [46], software professionals on the autism spectrum [26], programmers who use cannabis [16], software professionals who use psychoactive substances at work [33], and whistleblowers in the software industry [14]. Additionally, some publications treated software professionals as a hidden population in a broader sense [11, 28, 29, 45], employing this approach as a strategy to mitigate sampling bias [4].

A noticeable trend across these studies is the focus on underrepresented groups within software engineering as hidden populations. For instance, LGBTQIA+ software professionals constitute a minority group in the industry, with transgender software professionals facing significant marginalization. Moreover, whistleblowing mechanisms, often utilized to report harassment and misconduct, address issues commonly encountered by women in software engineering, which is another underrepresented group in the field. Overall, the upsurge in research concerning equity, diversity, and inclusion within software engineering sheds light on more hidden populations. Consequently, in empirical software engineering, there is a growing necessity to anticipate and manage the distinct challenges associated with studying these groups.

## 3 RESEARCHING HIDDEN POPULATIONS IN SOFTWARE ENGINEERING

This paper reports our experience conducting four studies involving two hidden populations in software engineering: LGBTQIA+ individuals in general [1] [12], and TGNC[2] individuals (in particular) [13, 19, 34]. In [12], we sampled LGBTQIA+ individuals for a grounded theory study, aiming to explore how remote and hybrid work is affecting these individuals. In [13], we were interested in understanding the career in software engineering from the perspective of transgender software professionals. In [34], we explored the needs and challenges of transgender individuals in hackatons. In [19], we worked with social open innovation in a Non-Governmental Organization (NGO) targeting people living with HIV and AIDS, whose main audience is mostly composed of young people, homosexual men, trans women, and sex workers. We also include reports from an additional study that was focused on investigating the trajectory of transgender and non-binary software professionals from academia into the software industry. However, it was never concluded due to challenges we could not overcome.

In general, all these studies shared a common goal: to obtain insights from these individuals and devise strategies to enhance diversity and inclusion in software engineering. However, conducting research involving these hidden populations presented challenges before, during, and after the studies. With this paper, we intend to share these experiences to aid fellow software researchers grappling with similar challenges.

### 3.1 Challenges

While researching underrepresented groups in software engineering, we faced challenges before—during the planning phase, during—while conducting the research, and after—in the reporting stage (Table 1). We believe our experience can support other software researchers dealing with hidden populations.

*3.1.1 Identifying participants who belong to the hidden population and sampling from within these individuals pose one of the greatest challenges in this context.*

Before initiating the study and outlining our approach to identifying individuals within the hidden population for data collection, we encountered issues commonly documented in the literature. Firstly, identifying suitable participants posed a challenge, as many software professionals from underrepresented groups might choose not to be identified. This hesitancy often stems from the pervasive lack of diversity in the software industry, prompting various hidden populations to conceal their background due to context stigma.

Nonetheless, issues in identifying individuals within a hidden population are often just the starting point. The sampling strategy must be all-encompassing, accommodating, and meticulously planned to capture a broad participant base. Implementing a probability sample becomes nearly impossible, as reported in other fields. For instance, have you ever considered *how many transgender software professionals you know? How can you randomize participant selection when there are not even many individuals who can participate in the study?* Hence, the sampling challenge extends beyond impractical randomization, as multiple non-probability techniques may still result in a small sample size, which is related to the subsequent challenge.

---

[1]Lesbian, Gay, Bisexual, Transgender, Queer/Questioning, Intersex, Asexual/Allies, and the plus sign is meant to cover anyone else who's not included [20]
[2]Transgender and gender non-conforming (TGNC) is an inclusive term that recognizes the diversity of gender experiences and identities beyond the binary perspective (male/female) and with examples of gender expressions including gender fluid, gender neutral [17].

**Table 1: Challenges in Investigating Hidden Populations in Software Engineering**

| Discrimination | Issue | |
|---|---|---|
| Study Planning | Identifying Participants | Defining and estimating hidden populations within software engineering, much like in diverse research fields, presents significant challenges. It is noteworthy that underrepresented groups in the software industry commonly constitute a portion of these hidden populations. |
| | Sampling | Non-probability sampling strategies are typically not viable for recruiting participants from hidden populations in software engineering. Sampling methods that employ a more interpersonal approach seem to be more effective in such recruitment endeavors. |
| Study Execution | Engaging Participants | Hidden populations in software engineering often comprise participants from underrepresented groups. To sustain their engagement and prevent withdrawal from the research, these participants need to clearly perceive the study's benefits to their communities. |
| | Authors Involvement | Belonging to or having close connections with a hidden population facilitates the identification and recruitment of participants. However, authors delving into sensitive topics within a population they belong to may unavoidably trigger psychological challenges and potentially affect the study. |
| Study Report | Paper Review | Convincing reviewers about the limitations of researching hidden populations in software engineering, while also advocating for the importance of investigating such populations, remains a significant challenge in this context. |

*3.1.2 The success of effective data collection depends on overcoming the challenge of engaging participants.*

Individuals from underrepresented groups in software engineering, belonging to hidden populations, tend to be selective when considering participation in studies. They can feel overwhelmed and tired of discussing sensitive topics repeatedly, which may trigger bad memories. Being part of an underrepresented group often involves regularly addressing sensitive issues like their pronouns, team dynamics, navigating toxic environments, facing implicit discrimination, and advocating for their rights within their job. Consequently, discussing these matters for research purposes might be the last thing they wish to engage with at the end of their day.

Related to this, a potential issue might happen during the execution of the study: individuals from hidden populations in software engineering might withdraw from research if they understand that the outcomes will not contribute to their or their group's welfare. In our experience, some participants initially agreed to join the study but later withdrew their participation. Others participated in the initial phase but opted out of subsequent sections. We observed that engaging participants from these hidden populations directly relies on demonstrating the research's relevance and the impact of the results on their groups. Although one may argue about sample size over 20 or even more participants in qualitative studies to reach data saturation, in some cases – especially with homogeneous populations – saturation is reachable in a narrow range of interviews (9–17) [24]. This would typically be the case in studies with hidden populations (e.g., TGNC people) which represent a group that is difficult to recruit participants. Thus, interviews must be more about depth than quantity.

*3.1.3 Authors who belong to the hidden population might opt to withdraw their involvement in the study.* Feeling overwhelmed and withdrawing from the study might not be limited to participants alone. In our experience investigating underrepresented groups in software engineering, we encountered a situation where the lead researcher decided to leave the study. During one of our studies mentioned above, the research topic evoked negative emotions in our co-authors while they were collecting data, recalling their own challenging experiences while they were interviewing participants from a similar background within the software industry. Consequently, this led them to request discontinuation of the study.

*3.1.4 Reporting the findings might pose a major challenge to authors due to the lack of awareness about the limitations of findings obtained from hidden populations among reviewers.*

Following the completion of research, reporting findings about hidden populations can pose challenges. While recruiting participants can be difficult, conference and journal reviewers may not fully grasp this perspective. Reviewers commonly prioritize sample size over result significance. They may question sample representativeness and undermine research findings if not all potential participants from the hidden population are reached. For example, in our journey aiming to enhance LGBTQIA+ inclusivity in the software industry, reviewers frequently critique the sample's imbalance in mainly including gays and lesbians while lacking representation from other groups within the community, such as bisexuals, queers, or asexuals. However, it is crucial to ponder: *How can these individuals be included if they opt not to disclose their sexuality?*

Therefore, after extensive efforts in identifying the hidden population, recruiting and retaining participants, and conducting data collection and analysis, a final challenge lies in convincing reviewers that increasing the visibility of this population through publications is a key strategy to encourage more individuals to engage in future research, which will gradually improve sample representativeness over time. In this particular context, embracing this progressive approach is more prudent than waiting for the optimal sample size before publishing studies because such a delay might hinder the investigation of important aspects of software development through the lenses of equity, diversity, and inclusion.

## 3.2 Lessons Learned

In our journey of studying hidden populations in software engineering, particularly software professionals from underrepresented groups, we navigated the aforementioned challenges mostly by combining or adjusting well-established methodologies. This enabled us to effectively collect data, perform analysis, and derive relevant insights aimed at improving equity, diversity, and inclusion in software engineering. Some key takeaways are cited below.

Nonetheless, it is essential to underscore that not all challenges encountered in our research were readily resolvable. Specifically, the withdrawal of one of our co-authors from the study due to their personal affiliation with the sensitive subject matter remained an unresolved issue. Furthermore, the reception of certain reviewers towards the constraints of studies conducted within hidden populations, in particular underrepresented groups, needs thoughtful consideration within the broader software engineering community.

*3.2.1 Purposive sampling helps to make the first contact with individuals within a hidden population, but snowballing sampling is more effective in reaching out to and retaining participants.*

In most of our studies, we employed communication channels within prominent software companies and online social communities to publicize our research and collect data through questionnaires, which enabled us to initiate contact with the hidden population. Through these questionnaires, we successfully engaged participants who preferred not to disclose their affiliation with the population. Upon concluding the questionnaire, we underscored the importance of our findings in potentially enhancing industry practices affecting this population. Additionally, we invited participants to join a subsequent study primarily based on interviews.

Although the number of participants willing to be interviewed remained consistently small, it initiated the referral-chain sampling (snowballing) process. During interviews, we encouraged participants to refer or nominate peers who could answer the questionnaires. As a result, we observed an increase in completed questionnaires post-interviews. Additionally, we have evidence of certain participants joining the study and accepting to participate in interviews only after consulting peers within the population.

> "I talked to [person's name], and he told me he knew you from when you worked at [company's name], that I could participate in our interview with no restrictions." [P03 during the interview to [13]]

*3.2.2 Convenience sampling is a useful initial strategy, especially when authors are part of the hidden population. However, to mitigate bias, transitioning from convenience sampling to employing theoretical sampling can yield more effective outcomes.*

As indicated in the literature, accessing hidden populations poses challenges as they tend to be more receptive to individuals within their community. In our experiences with underrepresented groups in software engineering, our studies were led by authors who were either part of or closely related to the population under study. This strategy facilitated participant identification through convenience sampling, for instance, an LGBTQIA+ researcher is likely to have connections with LGBTQIA+ software industry professionals. However, this approach can intensify biases inherent in convenience sampling, including selection bias, location bias, and response bias.

To address these biases, we employed convenience sampling to initiate implementing a theoretical sampling strategy. Theoretical sampling, frequently used in qualitative research methods such as grounded theory, involves selecting elements for study based on emerging findings. It allows researchers to determine their subsequent focus based on the ongoing analysis. In our case, we utilized convenience sampling to engage with familiar individuals from the population, aiming to gather insights from other participants with specific experiences or characteristics.

As an example, in [12], a participant emphasized that many software professionals who identify as LGBTQIA+ were relying on remote work as a means of isolating themselves to avoid discrimination in the workspace. To delve deeper into this emerging result, we engaged with other software professionals we were acquainted with (using convenience sampling) not to gather data, but to identify potential individuals who could provide us with further evidence to explore this scenario.

*3.2.3 Supplementary evidence on hidden populations might be available in the form of grey literature.*

The endeavor to gather comprehensive and varied data from hidden populations presents inherent challenges due to the secluded nature of these groups. Therefore, while engaging with participants from these populations formed a pivotal part of our research efforts, supplementing this primary data with evidence from grey literature became essential to capture at times to support a more comprehensive understanding of their experiences and viewpoints. Within the domain of hidden populations in software engineering, we observed that these groups often disseminate information through blogs, social media platforms, and developer forums. These sources can provide invaluable insights, opinions, and perspectives on various topics. Therefore, following a meticulous evaluation of the reliability and credibility of such data, we could discover its potential in shedding light on the social, cultural, or environmental influences that significantly shape their experiences within the software engineering topics under study.

*3.2.4 Building trust and fostering understanding among hidden populations is essential.* In our research conducted with an NGO serving transgender individuals [19], the NGO's coordinator disclosed a trust issue with academics. He observed that the transgender community served by his NGO often feels that researchers are primarily interested in extracting information, typically through interviews, without offering any tangible benefits or contributions in return.

This is somehow highlighted by Minalga et al. [32], who raise critical questions about the real beneficiary of research with transgender people, probing whether it serves the researchers' interests or offers real benefits to the transgender community. It calls for a reassessment of the research focus, urging a shift toward addressing more urgent and relevant issues faced by transgender individuals, especially those from marginalized groups like Black, Indigenous, and people of color. Additionally, the concerns highlight the need for ethical integrity in the research methodology, emphasizing the importance of avoiding potential harm or stigmatization of the transgender community. This underscores the necessity for research that is not only academically sound but also socially responsible and beneficial to the community it studies.

Particularly, in studies involving transgender communities, there are important guidelines that must be followed [41]. It is essential to understand the historical context of transgender communities, avoid past research mistakes, and maintain transparency throughout the research process, including the co-production of research questions, for instance, having TGNC team members in the research team. The use of inclusive and respectful language that acknowledges the community's diversity is crucial. Building trust and rapport while acknowledging the intersectionality of transgender identities with other social factors is vital for a comprehensive understanding. Furthermore, ensuring that research practices respect and do not harm transgender spaces and communities is key to successful participant engagement and effective data collection.

An example of statements that could not be captured without TGNC team members, one interviewee in [19] emphasized the "we" referring to herself and her interviewer, implicitly referring as "we transwomen":

> "**we** (transwomen) are very smart, **we** have a lot of ability to innovate."

### 3.3 Opportunities

Studies involving hidden populations in software engineering often lean towards employing qualitative research methods, as demonstrated in the studies outlined in Section 2. Qualitative approaches, such as interviews, focus groups, ethnography, and case studies, enable researchers to delve deeply into the nuanced experiences, perceptions, and behaviors of populations whose size is typically unknown, hard to estimate, and whose participants cannot be easily identified or sampled. However, the broader literature presents methods and strategies to conduct studies with hidden populations using a more quantitative perspective, which includes estimations, randomizations, and generalizations. These approaches offer research opportunities within empirical software engineering.

For instance, many studies in other disciplines have employed respondent-driven sampling to estimate the size of hidden populations, aiming for more generalizable results [7, 21, 23, 36]. Employing respondent-driven sampling involves selecting a group of individuals ("seeds") from the target population who are not yet in the sample via non-random means. These seeds are then tasked not with referring or nominating others but with recruiting individuals from their social network [22]. Subsequent waves of recruitment continue, and while the sample size expands, these waves are expected to reduce bias from highly connected participants, a potential issue in snowball sampling [4].

In fact, while respondent-driven sampling aims for a more representative sample from the hidden population by utilizing social connections, snowball sampling expands an initial sample through referrals from initial participants [4]. The underutilization of respondent-driven sampling in software engineering presents a significant opportunity for further exploration, both in estimating population sizes and its infrequent use as a sampling strategy. This gap provides a potential avenue for further research, particularly concerning hidden populations. Additionally, the literature encompasses various methodologies for estimating population sizes, such as Bayesian methods [38], state-space representation [39], and diverse mathematical models, which could be of interest in empirical software engineering studies.

In our quest to engage hidden populations in software engineering, we encountered location sampling, a method that centers on sampling from specific geographic areas or locations rather than solely focusing on individuals [9]. Location sampling encompasses physical spaces (e.g., events) and virtual spaces (e.g., social media groups or threads) [35]. While researching underrepresented groups in software engineering, we have yet to utilize this method. However, we recognize its potential usefulness for certain populations, depending on the investigated topic. For instance, social media channels used by LGBTQIA+ software professionals may serve as locations for sampling due to discussions on various work-related aspects, including whistleblowing. Similarly, events tailored for women in the software industry could be potential locations. Furthermore, the infrequent use of this method in software engineering studies points towards its potential for future investigations within empirical software engineering research.

Ultimately, we identify promising opportunities for discussions within the empirical software engineering community regarding the significance of evidence gathered from hidden populations using non-probability techniques like purposive or snowball sampling. Our community needs to recognize the real-world constraints that prevail in certain facets of software engineering, particularly concerning the fact that many hidden populations within our field consist of individuals from underrepresented groups. Gaining insights into these groups of software professionals is pivotal for cultivating a more diverse and inclusive software engineering. However, studies focusing on these populations might face inherent limitations. Hence, we must engage in ongoing discussions in other research domains [5, 8], reflecting on the balance between methodological rigor, the practical aspects of real-world settings, the characteristics of hard-to-reach populations, and the limitations of research methods. This critical reflection holds paramount significance for research in equity, diversity, and inclusion within software engineering and presents a significant opportunity to advance empirical software engineering practices toward mitigating bias and improving reviewing processes.

## 4 FINAL CONSIDERATIONS

Hidden populations in software engineering comprise groups of individuals that are difficult to identify and reach due to various reasons regarding their marginalized status, seclusion, or reluctance to participate in research. These groups typically include professionals from underrepresented in the software industry, such as women, LGBTQIA+ individuals, ethnic minorities, people with disabilities, and those encountering discrimination. Our experience researching these populations has revealed significant insights and lessons on empirical methods. For instance, challenges in this context encompass difficulties in identifying the population, sampling issues, participant and researcher withdrawal, and substantial criticism in reviews overlooking empirical method limitations in dealing with hidden populations. These challenges highlight the importance of discussions within the research community to foster a more inclusive research environment. Addressing research-related issues concerning hidden populations is crucial in understanding the unique challenges faced by underrepresented groups within the software industry. Gaining insights into their experiences and distinctive difficulties is key to resolving systemic issues. Consequently, resolving research challenges in this context plays a pivotal role in fostering an inclusive software engineering environment where individuals from diverse backgrounds can thrive and contribute significantly.

## ACKNOWLEDGMENTS

This work is partially supported by INES (www.ines.org.br), CNPq grant 465614/2014-0, FACEPE grants APQ-0399-1.03/17 and APQ/0388-1.03/14, CAPES grant 88887.136410/2017-00.